\documentclass[11pt]{article}
\usepackage{amsmath}
\usepackage{color}

\usepackage{amsfonts}
\usepackage{amssymb}
\usepackage{graphicx}
\usepackage{subfigure}

\begin{document}

\newtheorem{problem}{Problem}
\newtheorem{definition}{Definition}
\newtheorem{lemma}{Lemma}
\newtheorem{proposition}{Proposition}
\newtheorem{corollary}{Corollary}
\newtheorem{example}{Example}
\newtheorem{conjecture}{Conjecture}
\newtheorem{algorithm}{Algorithm}
\newtheorem{theorem}{Theorem}
\newtheorem{condition}{Condition}
\newtheorem{exercise}{Exercise}

\newcommand{\la}{{\lambda}}
\newcommand{\eps}{{\varepsilon}}
\newcommand {\bA} {\mbox{\boldmath $A$}}
\newcommand {\bX} {\mbox{\boldmath $X$}}
\newcommand {\bY} {\mbox{\boldmath $Y$}}
\newcommand {\bZ} {\mbox{\boldmath $Z$}}
\newcommand {\bfa} {\mbox{\boldmath $a$}}
\newcommand {\bx} {\mbox{\boldmath $x$}}
\newcommand {\by} {\mbox{\boldmath $y$}}
\newcommand {\bz} {\mbox{\boldmath $z$}}

\newcommand{\ls}[1]
   {\dimen0=\fontdimen6\the\font \lineskip=#1\dimen0
\advance\lineskip.5\fontdimen5\the\font \advance\lineskip-\dimen0
\lineskiplimit=.9\lineskip \baselineskip=\lineskip
\advance\baselineskip\dimen0 \normallineskip\lineskip
\normallineskiplimit\lineskiplimit \normalbaselineskip\baselineskip
\ignorespaces }

\title{A Note on the Compaction of long Training Sequences for Universal
Classification---A Non-Probabilistic Approach}
\author{Jacob Ziv\\
Department of Electrical Engineering\\
Technion--Israel Institute of Technology\\
Haifa 32000,  Israel}
\date{\it March 29, 2014}

\maketitle
\begin{abstract}
One of the  central problems in the classification of individual test
sequences (e.g. for genetic analysis), is that of checking  for the similarity of
sample test sequences of length $N$ letters as compared with a set of  much
longer training sequences. This is done by a set of classifiers for test
sequences of length $N$, where each of the classifiers is trained by a
corresponding individual training sequence so as to  minimize the  classification error rate of the
classifier when fed with the given corresponding individual training sequence.

It should be noted that the storage of long training sequences is considered
to be a serious bottleneck in the next generation sequencing for Genome analysis.

Some popular classification algorithms adopt a  probabilistic approach, by
assuming that the sequences are realizations of some variable-length Markov
processes or a hidden Markov process (HMM), thus enabling the embedding of the
training data onto a variable-length pruned Suffix-tree, the size of which is usually
linear in $N$, the length of the test sequence, rather than the much longer training sequence.

Despite the fact that it is not assumed here that the sequences are
realizations of probabilistic processes (an assumption that does not seem to
be fully justified when dealing with biological data), it is demonstrated that
``feature-based'' classifiers, where particular substrings (called ``features'' or ``markers'')
are sought in a set of ``big data'' training sequences, may be utilized to efficiently classify test sequences,
without adapting a probabilistic model. This is achieved by applying universal compaction of the training
data that is contained in a  set of $t$ (long) individual training sequences,
onto a compact suffix-tree the size of which (similar to the probabilistic case)
is linear in $N$, rather than in the length of the much longer training sequences,
regardless of how long they are, at the cost of only a vanishing increase in the empirical
misclassification errors of test sequences, relative to the individual long training sequence.
This justifies the  efficient compaction of the long training data onto a  suffix-tree training data base
 without relying on any probabilistic models.

Furthermore, it is demonstrated that when the long training sequences are
each compacted onto a pruned suffix tree, such the suffix trees intersect with each other,
it is possible to further compact these trees by merging all the trees
onto one tree, with a total number of leaves that may be much smaller than
the total number of leaves of all of the individual trees (relative compaction).
\end{abstract}
\section{Introduction}
Some  popular classification algorithms adopt a probabilistic
approach, by assuming that the sequences are realizations of
some variable-length Markov process or a hidden Markov process (HMM),
thus enabling the embedding of the training data onto a pruned variable-length
Suffix-tree, the size of which is usually linear in $N$, the length of
the test sequence, regardless of how long the training data is.(e.g., \cite{1},\cite{3},\cite{5},\cite{6}).

Despite the fact that the probabilistic approach is not necessarily theoretically
justified, it apparently led to good empirical classification results. and also  are shown to be asymptotically
classifiers  under the Markov process regime.

In this note it is demonstrated that despite the fact that  no probabilistic
model for the data is utilized and sequences are treated as individual
sequences, it still efficiently leads again to the enabling the embedding of
the
training data onto a variable-length Suffix-tree, the size of which is usually
linear in $N$, the length of the test sequence, at only a negligible
increase in the False-Negative empirical error rate.

It should be noted here that in the case of data compression of long
sequences, an alternative to the probabilistic approach was established by assuming that
the stream of data to be compressed is a non-probabilistic \textbf{individual} sequence.

The assumption that the compression is carried out via a universal Turing
machine led to the notion of Kolmogorov complexity, which is the best
asymptotic compression ratio that may be achieved for the individual sequence
by any computer.

A more conceptually restricted, but practical approach, was obtained by
replacing the universal Turing machine model by a  finite-state machine (FSM)
model or a finite block-length compression model (LZ) \cite{2},\cite{9},\cite{7}.
This approach led to an associated suffix-tree data base with $O(N')$ leaves, where $N'$
is the length of the sequence to be compressed and where all
leaves have about the same empirical probability of appearance.

It has also been demonstrated that the common probabilistic modeling approach
for prediction tasks may  be replaced by an individual sequence approach as well.

In this case too, organizing the data base in the form of a variable-length
suffix-tree (context-tree) with leaves that have about the same empirical
probability of appearance of suffixes, has led to efficient on-line prediction \cite{4}.

A similar approach is adapted here, by studying the performance of universal
classifications of an individual test sequence relative to a long individual training sequence.

In this note we study the classification of individual test-sequences relative
to a collection of substrings (\textbf{features}) that may be
embedded in one or more of $t$  individual, long training sequences.

Despite the fact that it is not assumed that the sequences are
realizations of a probabilistic process (an assumption that does not seem to
be fully justified when dealing with biological data), it is demonstrated that
it is possible to compact the $t$ training sequences
onto $t$ pruned suffix trees with typically only $O(N)$ leaves, at only a
small increase in the empirical False-Negative error rate, even if the $t$
suffix trees are only  slightly different from each other. It is possible to
find an optimal collection of $t$ feature sets $F_i \, ; \; i=1, 2, ..., t$
that will make the set of training sequences \textit{separable}, and further compact
the $t$ trees by merging the $t$  individual trees onto one tree with a
number of leaves that may be much smaller than the total number of leaves of the $t$ individual trees.

Furthermore, the generation of  the suffix-trees from the training sequences
is  \textit{universal}, since it does not depend on the specific feature set or a specific similarity function.

\section{Definitions and Derivations}
\label{sec:2}

Consider the classification of individual test sequences (e.g.
shot-gun reads for genetic analysis).
The test sequence is denoted by ${\bf X}=x_1,x_2,...,x_{N};x_i \in \bf A$ of
length $N$ letters, where $\bA$ is an alphabet of $A$ letters.
The purpose of classification is that of checking the similarity of a sample test sequence $\bX$ to an
extremely long (``big data'') sequence ${\bf Y}=y_1,y_2,...,y_{N'};y_i \in \bf A$ of length $N'>>N$.

It is necessary to decide whether the test sequence $ \bX =(x_1,x_2,...,x_N) $
is  more  similar to one out of $t$ very long individual training  sequences
$ \bY_i \, ; \; i=1,2,...,t$, or not similar to any of these training sequences.

This is achieved by a classifier $C(\bY_{i}) \, ; \; i=1,2,...,t$, utilizing
a corresponding ``similarity function'' $ S(F_i;\bX)\geq 0$, where $F_i$ is a ``feature set''
that consists of $f_i$ ``typical'' substrings (that are sought in $ \bY_i$),
denoted by $ \bZ_{i} (k) \, ; \; k=1,2,...,f_i $, where no substring in $F_i$ is a
prefix of any other substring in $F_i$.

It is assumed that $S(F_i;\bX)= 0$, if no substring in $\bX$ is an element of the feature set $F_i$.
It is also assumed that the cardinality of $F_i$ is no larger than $N$
(this is a typical  assumption in the probabilistic-based classification approach).
A test sequence $\bX$ is declared to be \textbf{more} similar to a training
sequence $\bY_i$ than to another training sequence $\bY_j$ if $S(F_i;\bX)>S(F_j;\bX)$.

Consider, for example a similarity function that is associated with a modified
version of the Average Common Length (ACL) classification algorithm \cite{6}.

Consider a tree that  consists of  $f_i$ leaves that correspond to all the elements of $F_i$.
Denote the length of an element ${\bf Z}_{i}(k);k=1,2,...,f_i$ of  the feature set $F_i$, by $L_i(k)$.
Let $\bX_i^j=x_i,x_{i+1},...,x_{j}$.

Let $$L(\bY_i)  =\displaystyle\sum\limits_{k=1}^{f_i}{\frac {1}{f_i}}L_i(k) \, .$$
Let  $\delta (\bX_1;\bX_2)=1$ if $\bX_1=\bX_2$, else $\delta
(\bX_1;\bX_2)=0$ and let
$$L(\bX
|\bY_i)=\frac{1}{N-1}\sum_{k=1}^{f_i}\sum_{j=1}^{N-L_i(k)+1}{\delta}(\bX_j^{j+L_i(k)-1};\bZ_i(k))L_i(k)
\, .$$
Let the similarity function be defined by:
$$ S(F_i;\bX)=\frac {L(\bX |\bY_i)}{L(\bY_i)} \, .$$
It follows from [6] that this similarity function leads to am asymptotically-optimal classifier under 
a finite memory Markov process regime.

In some cases of interest the feature set is a set of substrings that serve
as genetic ``markers'' that typify the training sequence $\bY_i$,
and the classification is based on the appearance of all or some of these
markers as substrings in the test sequence $\bX$.

\smallskip
\noindent
\textbf{Classification Error Rate and Training}

Assume that the length of the training sequence, N' is much larger than the length of the test sequence, $N$.

Given a similarity function $S(F_i;\bX)$, and a set of $t$ training sequences,
the classifiers  $C(\bY_{i}) \, ; \; i=1,2,...,t$ are trained as follows:

Denote by $\bY (k,N)$ a substring of $\bY$ of length $N$,$\bY(k,N)=y_{k}, y_{k+1},..., y_{k+N-1};k=1, 2, ..., N'-N+1$.

Let $\bY_i(k,N)$ satisfy $S(F_i;\bY_{i}(k,N))>0$, and define $q(i, \bY_{j}(k,N))=1$
if for some  other distinct subsequence of length $N$ that is not an element of the feature set $F_i$,
$\bY_{i}(k',N);k'\neq k$  in $\bY_i$, $S(F_i;\bY_{j}(k',N))\geq S(F_i;\bY_{i}(k,N))$.

Else, set $q(i, \bY_{j}(k,N))=0$.

Let ${\delta}(fp|i)=\frac{1}{N'-N+1}\sum_{k=1}^{N'-N+1}q(i,( \bY_{j}(k,N))$.

Let ${\delta}(i|i)$ denote the number of subsequences of length $N$
in $\bY_i$ that is not an element of the feature set $F_i$ , for which $S(F_i; \bY_{i}(k,N))>0$.

Then $\delta(fp|i)$ is named: the (False-positive) \textit{classification error rate}.

Also, let
${\mu}(fn|i)=\frac{1}{N'-N+1}\sum_{k=1}^{N'-N+1}\delta(\bY_{j}(k,N))\in F_i;S(F_i;\bY'_{j}(k,N))>0)=0$,
then $\frac{{\mu}(fn|i)}{\delta(i|i)}$ is named the (False-negative)
\textit{classification error rate} and is
the empirical frequency of substrings in $\bY_i$, which are elements of the feature set $F_i$
but are not recognized by the classifier.

A classifier $C(\bY_i;\bX)$  is \textit{acceptable} and well trained relative
to the training sequence $\bY_i$, if it achieves
${\delta}(i|i)\geq \delta$,  ${\delta}(fn|i)\leq {\epsilon}{\delta} $  and ${\mu}(fp|i)\leq (\delta)(\mu)$,
where $\delta$, $\epsilon$ and $\mu$ are some preset parameters.

\section{Universal Compaction of a Collection of Individual Training Sequences}
\label{sec:3}

Next, a \textit{universal} compaction of individual training sequences is proposed
that
does not depend on any a-priori information about the set of training sequences, and
is efficient and linear in $N$ without deteriorating the performance of any
classification algorithm for test sequences
of length $N$ relative to long  individual training sequences of length $N'>>N$, except for a small prescribed
additional False-negative empirical classification error-rate.
\begin{enumerate}
\item
For every $\bY_i$ generate a tree $T_i$ that consists of all
substrings $\bY_i(k,l) \,; \; k=1, 2,..., N'-l+1; l=1, 2,..., N$,
for which $p_{i}^{*}(\bY_i(k,l))\geq \frac{{\mu}\delta}{f}$, where $f\leq N$
is the largest value of $f_i \, ; \;i=1, 2, ...,t$,
and store the empirical probabilities of all the leaves and nodes
of the tree $T_i$.

It follows by construction  that the number of leaves of $T_i$ is no larger than $\frac{N}{(\mu)\delta}$.
\item
Merge the $t$ trees onto one tree $T$.
\end{enumerate}

Observe that some collections  of $t$ training sequences may yield a
merged tree where the number of leaves is much smaller than the total number of leaves of all the trees $T_i \, ; \; i-1, \, 2,... $.

Similar to the probabilistically-motivated classification algorithms that are cited above and that utilize as a training data-base one that is
generated by efficiently compacting long training sequences onto a suffix-tree together with its associated  probabilities,
 the compaction  procedure that is described above also  utilizes a similar compaction and
 favorably compares with any classifier $C(\bY_i,\bX)$ for a given feature set,
 and the set of the associated $\it empirical$ probabilities of the different  features in $\bY_i$
 (e.g. the version of the ACL algorithm that appears above), at the cost of a slight increase in the
 False-negative empirical error rate by at most $\frac{\mu}{\delta}$:

The merged pruned suffix tree
$T$ that is thus generated from the $t$ training sequences,
together with the corresponding collection of the empirical probabilities as defined in
the compaction procedure above, universally serves as sufficient training statistics for
classifying test sequences of length $N$  at the cost of an increase of at
most $\frac{\mu}{\delta}$.

 This justifies the  efficient compaction of the long training data onto a  suffix-tree training data base
 without relying on any probabilistic models.
\

  \section*{Acknowledgment}

Helpful comments by Alberto Apostolico, Nahum Shimkin, Neri Merhav and Michal Ziv-Ukelson are acknowledged with thanks.




\begin{thebibliography}{10}

\bibitem{1}
G. Bejerano and G. Yona, ``Variations on probabilistic suffix trees---a
new tool for statistical modeling and prediction of protein families,
Bioinformatics, 17(1): 23--43, 2001.

\bibitem{2}
J. Ziv and A. Lempel, ``A Universal Algorithm for Sequential Data
Compression'', IEEE Trans. Inf. Theory, vol.~IT--21, 1967, vol.~23: 337--343, 1977.

\bibitem{3}
R. Giancarlo, D. Scaturro and F. Utro, ``Textual Data Compression In Computational Biology:
A synopsis'', Bioinformatics, 2(13): 1575--1586, 2009.

\bibitem{4}
J. Ziv and N. Merhav, ``On Context-Tree Prediction of Individual Sequences,''
IEEE Trans. Inf. Theory, pp.~1860--1866, 2007.

\bibitem{5}
G. Reinert, D. Chew, F. Sun and M. S. Waterman, ``Alignment-Free
Sequence Comparison (I): Statistics and Power,'' J. Comput. Biol., 16(12): 16151634, 2009.

\bibitem{6}
I. Ulitsky, D. Burstein, T. Tuller  and B. Chor, ``The average common
substring approach to phylogenomic reconstruction,'' J. Comput. Biol., 13(2): 336--350, 2006.

\bibitem{7}
F. M. J. Willems, Y. M. Shtarkov and T. J. Tjalkens, ``The context-tree
weighting method: basic properties'', IEEE Trans. Inf. Theory, 41(3): 653--664, 1995.

\bibitem{8}
J. Ziv and N. Merhav, ``A measure of relative entropy between individual
sequences with application to universal classification,'' IEEE Trans. Inf. Theory, 39(4): 1270--1279, 1993.

\bibitem{9}
J. Ziv, ``On Finite-memory Universal Data-compression and Classification of Individual
Sequences'', IEEE Trans.  Inf. Theory, 54(4):1626--1636, 2008.

\bibitem{10}
S. M. Brown, Bioinformatics NYU School of Medicine---PowerPoint PPT Presentation.
\end{thebibliography}
%



\end{document}